\documentstyle[prl,aps,epsfig,multicol]{revtex}
\begin{document}
\def\be{\begin{equation}}
\def\ee{\end{equation}}
\def\bearr{\begin{eqnarray}}
\def\eearr{\end{eqnarray}}
\def\tc{$T_c~$}
\def\tcl{$T_c^{1*}~$}
\def\c2{ CuO$_2~$}
\def\ruo{ RuO$_2~$}
\def\lsco{LSCO~}
\def\bi{bI-2201~}
\def\tl{Tl-2201~}
\def\hg{Hg-1201~}
\def\sro{$Sr_2 Ru O_4$~}
\def\rc{$RuSr_2Gd Cu_2 O_8$~}
\def\mgb{$MgB_2$~}
\def\pz{$p_z$~}
\def\ppi{$p\pi$~}
\def\sqo{$S(q,\omega)$~}
\def\tperp{$t_{\perp}$~}
\draft
\title{Predicting a Gapless Spin-1 Neutral Collective Mode 
branch for Graphite} 

\author{G. Baskaran$^{1}$, S. A. Jafari$^{1,2}$}

\address{$^{1}$ Institute of Mathematical Sciences, Madras 600 113, India\\
$^{2}$Department of Physics, Sharif University of Technology,
P. O. Box: 11369-9161, Tehran, Iran}
\date{\today}
\maketitle
\begin{abstract}
Using the standard tight binding model of 2d graphite with  
short range electron repulsion, we find a gapless  spin-1, neutral 
collective mode branch {\em below the particle-hole continuum} with 
energy vanishing linearly with momenta at the $\Gamma$ and $K$ 
points in the BZ. This spin-1 mode has a wide energy dispersion, 
$0$ to $\sim 2~eV$ and is not Landau damped.  
The `Dirac cone spectrum' of electrons at the chemical potential of 
graphite generates our collective mode; so we call this
`spin-1 zero sound' of the `Dirac sea'. Epithermal neutron scattering 
experiments, where graphite single crystals are often used 
as analyzers (an opportunity for `self-analysis'!), and spin polarized 
electron energy loss spectroscopy (SPEELS) can be used to confirm and
study our collective mode.
 
\end{abstract}

\begin{multicols}{2}[] 
Graphite is an important system in condensed matter science and 
technology; in carbon research its role is fundamental. Its electrical 
and magnetic properties have been investigated for decades both 
experimentally and theoretically\cite{dress1}. 
It is one of the simplest of quasi two dimensional 
zero gap semiconductors/semi metals. Intercalated graphites offer 
many phases of condensed matter including superconductivity.
Other important systems 
such as Bucky balls, carbon nano tubes\cite{dress2}
and some form of amorphous carbon derive many of their novel 
properties from their underlying `graphite character'. Any newer 
understanding of graphite is likely to have a wider impact. 

The aim of the present letter is to 
predict a simple but important property of graphite that calls for 
re-examination of some of the low energy electrical and magnetic 
properties of graphite. We find that
graphite possesses a new, unsuspected gapless branch of a spin-1 
and charge neutral collective mode. This branch lies below the 
electron-hole continuum (figure 1); 
its energy vanishes linearly with momenta as
$ \hbar\omega_s \approx \hbar v_F q ( 1 - \alpha q^2) $ 
about three points ($\Gamma$,$K$ $K'$) in the BZ (figure 1).

Since graphite interpolates metals and insulators, our collective mode
can be viewed both from metallic and insulating stand point.
In paramagnetic metals `zero sound' is a Fermi surface
collective mode\cite{pines}. The `charge' as well as the 
`spin' of a Fermi sea can undergo independent oscillations.
The charge oscillation becomes a high energy branch, the plasmon, 
because of the long range coulomb interaction; plasmons in graphite
has been studied in great detail in the past\cite{plasmon.gr}. 
The electron-electron interactions in normal metals do not usually 
manage to develop a low energy spin collective mode branch because of 
the nature of the particle-hole spectrum.
However, {\em the particle-hole spectrum of  2d graphite with 
a `window' (figure 2) provides an unique
opportunity for a  spin-1 
collective mode branch to emerge in the entire BZ}. From this point of view
our spin-1 collective mode is a `spin-1 zero sound'(SZS) 
of a 2+1 dimensional `massless Dirac sea', rather than a `Fermi sea'.

From an insulator point of view our collective mode is a spin-triplet 
exciton branch. Triplet excitons are well known in insulators, 
semiconductors and $p\pi$ bonded planar organic molecules; however, 
they usually have a finite energy gap, except when there are magnetic
instabilities.  

Our spin-1 collective mode may be thought of as a manifestation of 
Pauling's\cite{pauling} RVB state of graphite: 
the spin-1 quanta is a delocalized triplet bond in a sea of resonating 
singlets. The gaplessness makes it a `long range RVB' rather than
Pauling's short range RVB. 
Later we will present an argument to suggest that at low energies the 
neutral spin-1 excitation might undergo quantum number fractionization 
into two spin-$\frac{1}{2}$ spinons. 

Existence of our gapless spin-1 collective mode branch should influence 
the spin part of the magnetic susceptibility, rather than the orbital part,
which for graphite is diamagnetic, large and anisotropic.
Study of spin susceptibility by ESR, NMR and inelastic neutron 
scattering are good probes to detect the low energy part of our 
collective modes over a limited energy up to $\sim 50~meV$.  
A recent observation of `large internal fields' in oriented pyrolitic 
graphite by Kopelevich and collaborators\cite{kop1}, in their ESR 
studies could be due to our low energy spin-1 collective modes around 
the $\Gamma$ point in the BZ. The low energy  collective modes also 
contribute to specific heat and thermal conductivity over a wide 
temperature range. Our mode could be probed
over a large energy range,  by epithermal neutrons and spin polarized electron 
energy loss spectroscopy (SPEELS)\cite{eels}. In view of a wide energy 
scale associated with the collective modes,
probes such as two magnon Raman scattering, ARPES, STM and 
spin valves\cite{spinvalve} should also be tried. 

Importance of electron-electron interaction in graphite\cite{paco,dimo}
and related systems\cite{dunghai,mathew,harigaya} has been realized recently 
and it has lead to several interesting studies and predictions.
2d cuprates with Dirac cone spectrum has been studied in the context of 
AFM order in the Mott insulating RVB-flux phase, for spin-1 goldstone 
modes\cite{muthu} and d-wave superconducting phases, for spin-1 
collective modes\cite{palee}.

Real graphite is a layered semimetal - stacked layers of honeycomb
lattice of carbon atoms. We have one \pz orbital as the relevant valence 
orbital and one electron per carbon atom.  The \ppi bond 
produces a filled valence band and an empty conduction band with 
vanishing band gaps at two $K$ points in the BZ.  The coupling between 
graphite layers is van der Waals like. 
However, a small `coherent' interlayer hopping has been invoked 
to explain the presence of small electron and hole tubes and
 pockets,
(with $10^{-4}$ carriers per carbon atom, i.e., a Fermi energy
$\epsilon_F \sim 100 - 200 K $), responsible for the 
semi metallic character of graphite. 

We start with a 2-dimensional Hubbard model for graphite, which 
captures the physics of low energy spin dynamics. The Hamiltonian is: 

\begin{eqnarray*}
      H=-t\sum_{\langle i,j\rangle,\sigma}{
(c^{\dagger}_{i,\sigma}c^{}_{j,\sigma}+h.c)}+
        U\sum_{i}{n_{i\uparrow}n_{i\downarrow}}
\end{eqnarray*}

Here $t \sim 2.5~eV$ is the nearest neighbor hopping matrix 
element. While the bare atomic U is of the order of $8~eV$, 
the effective renormalized U can be of the order of $ 3 - 4~eV$. 
We will keep U as a parameter to be fixed by experiment.  

The dispersion relation for the \ppi-bands is:
\begin{equation}
\varepsilon_{\bf k} = \pm t\sqrt{ 1 +
4\cos \frac{{\sqrt 3}k_xa}{2}\cos\frac{k_ya}{2}+
4\cos^2\frac{k_ya}{2} }
\end{equation}
with vanishing gaps at the two $K$ points in the BZ (figure 1).
The particle-hole continuum of excitations is shown in figure 2.  The 
`Dirac cone single particle spectrum' at the $\Gamma$ and $K$ points 
makes the particle hole continuum very different 
from that of a free Fermi gas, or systems with extended Fermi surface. 
In contrast to figure 3, the particle-hole spectrum of a 2d Fermi 
liquid, our spectrum has a `window'. {\em The `window'
is characteristic of a 1d particle-hole spectrum}. 
In the Hubbard model two particles with opposite spins at a given 
site repel with an energy U. This means an attraction for up spin 
particle and down spin hole; or an attraction in the spin triplet
channel for a particle-hole pair. A spin triplet particle-hole pair 
could form a bound state, provided there is sufficient phase space 
for the attractive scattering. We find one spin-1  bound state for every
center of mass momentum of the particle-hole pair. In particular
an effective 1d character of phase space also makes the collective
mode energy vanish linearly with momenta around the three points:
$\Gamma$ and $K$'s.
\begin{figure}[h]
\epsfxsize 7cm
\centerline {\epsfbox{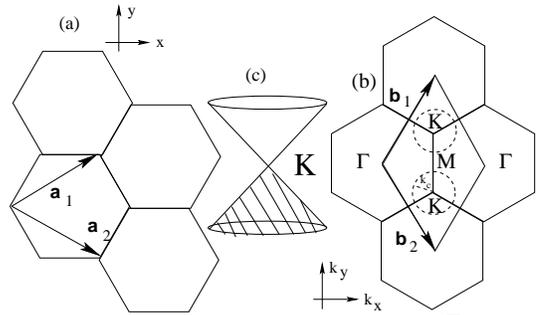}}
\caption{(a) Honeycomb lattice; ${\bf a}_{1,2} = 
\frac{a}{2}({\sqrt 3},\pm 1)$ (b) the Brillouin Zone;
${\bf b}_{1,2} = \frac{2\pi}{a}(\frac{1}{\sqrt 3},\pm 1)$ 
(c) Dirac cone spectrum at a $K$ point.} 
\end{figure}
The collective mode that we are after are obtained as the poles of 
the particle-hole response function in the spin triplet channel. 
We will focus on the zero temperature case.
The magnetic response function within the RPA (particle-hole ladder 
summation) is given by:
\begin{equation}              
\chi({\bf q},\omega)
=\frac{\chi^{0}({\bf q},\omega)}
{1-\upsilon({\bf q})\chi^{0}({\bf q},\omega)}
\end{equation} 

For Hubbard type on site repulsion , $\upsilon({\bf q})= U$
and the {\it free particle} susceptibility is :

\begin{equation} 
        \chi^{0}({\bf q},\omega)=\frac{1}{N}\sum_{{\bf k}}{\frac
        {f_{{\bf k}+{\bf q}}-f_{{\bf k}}}{\omega -(\varepsilon_{{\bf k}+
        {\bf q}}-\varepsilon_{{\bf k}})}}
\end{equation}
Here $f$'s are the Fermi distribution functions.
We have evaluated the RPA response function numerically and
found the collective mode branch in the entire BZ, below the 
particle-hole continuum. However, it
is instructive to linearize the electron and hole dispersion for
low energies,  a Dirac cone approximation\cite{paco}, and get an analytical  
handle.  We linearize the dispersion around $K$ and $K'$
and replace the BZ by two circles of radii $k_c$
(figure 1): 
\begin{equation}
\varepsilon_{{\bf k}} =  \pm v_{F}\mid{\bf k}\mid ~~
{\mbox for} k < k_c 
\end{equation}
where $v_{F}=\frac{\sqrt{3}}{2}t$ and $N$ is the number of unit cells.  
In our linearization scheme, in equation 4 the summation is over the 
two circular patches (figure 1b).

For a finite range of $q$ and $\omega$, $Im~\chi^0(\omega,{\bf q})$ 
can be evaluated exactly\cite{paco}:
\bearr
 Im~\chi^{0}({\bf q},\omega)=\frac{1}{16 v_{F}^{2}}
        \frac{2\omega^{2}-(q v_{F})^2}{\sqrt{\omega^{2}-(q v_{F})^2 }}
\sim 
  \frac{1}{16\sqrt{2 v_{F}}}\frac{q^{3/2}}{\sqrt{\omega-q v_F}}~~,
\nonumber
\eearr
with a square root divergence at the edge of the particle-hole continuum
in ($\omega, {\bf q}$) space. 
This expression has the same form as density of states of a  
particle in $1D$ (with energy measured from $v_Fq$).  Note that in fact 
$Im~\chi^{0}(q, \omega)=\pi \rho_q(\omega),$ where
$\rho_q(\omega)$ is free particle-hole pair $DOS$ for a fixed center of
mass momentum $q$. {\em That is, the particle-hole pair has a
phase space for scattering which is effectively one dimensional}. 
Thus we have a particle-hole bound state in the spin triplet channel
for arbitrarily small U. However, we also have a prefactor $q^{3/2}$, 
that scales the density of states. This together with  
\begin{figure}[h]
\epsfxsize 7cm
\centerline {\epsfbox{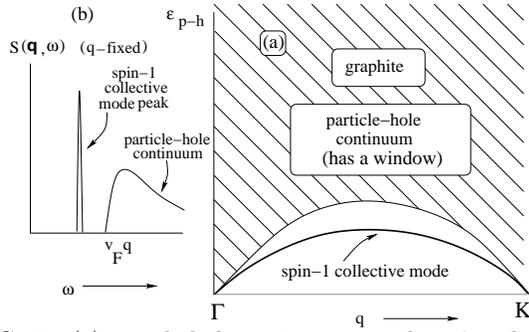}}
\caption{(a)particle-hole continuum with a `window' for
graphite. (b) $S({\bf q},\omega)$  for $q > q_c (\sim \frac{1}{50}
\frac{\pi}{a})$} 
\end{figure}
the square root divergence of the density of states at the
bottom of the particle-hole continuum  gives us a bound state for 
every q as $q \rightarrow 0$, with the binding energy vanishing 
as $\alpha q^3$, as shown below.
The square root divergence has the following phase space interpretation. 
The constant energy $(\hbar\omega)$ contour of a 
particle-hole pair of a given total momentum $q$ defines an ellipse
in k-space: $\omega = v_{F}(\mid{\bf k}+{\bf q}\mid+\mid{\bf k}\mid)$.
In our convention, the points on the ellipse denote the 
momentum co-ordinates of the electron of the electron-hole pair.  
As the energy of the particle-hole pair approaches the bottom
of the continuum, i.e., $\epsilon_{p-h} \rightarrow v_F q$, the minor 
axes of the ellipses become smaller and smaller and the elliptic contours
degenerate into parallel line segments of effective length 
$\sim q^{\frac{3}{2}}$. The asymptotic equi-spacing of these line 
segments leads to an effective one-dimensionality and the resulting
square root divergence.

According to $(1)$, the collective mode in $magnetic$ channel is the 
solution of:

\begin{eqnarray*}
        1-U\chi^{0}({\bf q},\omega)=0
\end{eqnarray*}
or equivalently,$ Im~\chi^{0}({\bf q},\omega)=0 ~~~{\mbox and}~~~ 
Re~\chi^{0}({\bf q},\omega)=\frac{1}{U}$.
The asymptotic expression for $ Re~\chi^0({\bf q},\omega)$ is found
to be

\begin{eqnarray*}
 Re~\chi^0({\bf q},\omega) \approx \frac{1}{4\pi^{2} v_F}
({k_c} + \frac{\sqrt{2}}{\sqrt{1-z}} \arctan {(\frac{\sqrt{2}}{\sqrt{1-z}})})
\end{eqnarray*}
where  $z\equiv \frac{\omega}{q v_{F}}$. Using the above expression
we obtain the following dispersion relation for the collective 
mode:
\be
	\omega  =  q v_F-\frac{q^3}{32\pi^2 v_F
	(\frac{1}{U}-\frac{k_c}{4\pi v_F})^2} \equiv q v_F - E_B(q)
\nonumber
\ee
as $\omega \rightarrow q \rightarrow 0$. Here $E_B(q)$ is the 
{\it binding} energy of the particle-hole pair of momentum $q$
around the $\Gamma$ point. The binding energy around the $K$ points 
is roughly half of this.

We mentioned earlier that our collective mode is a `magnetic zero
sound'. While magnetic zero sound are difficult to get in normal
metals, graphite manages to get it in the entire BZ because of the 
window in the  particle-hole spectrum (figure 2). 
\begin{figure}[h]
\epsfxsize 7cm
\centerline {\epsfbox{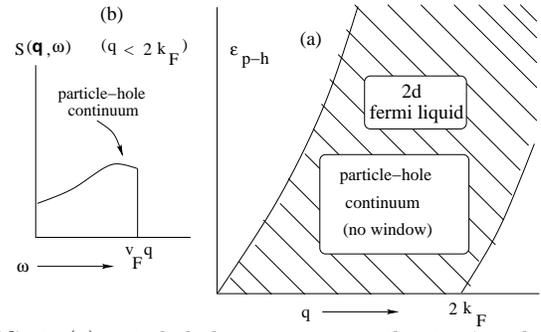}}
\caption{(a)particle-hole continuum without a `window' for
a 2d fermi gas. (b) $S({\bf q},\omega)$  for $q < 2k_F$.} 
\end{figure}
Having established the existence of a gapless spin-1 collective mode branch
within Hubbard model and the RPA approximation, we will discuss
whether the semi-metallic screened interaction of 3d stacked layers
will affect our result.  As mentioned earlier, 
in tight binding situation like ours, the spin physics is mostly 
captured by the short range part of the 
repulsion among the electrons. We have numerically studied the 
response function for a more realistic intra layer interaction 
namely the screened coulomb 
\noindent
interaction (including interlayer 
scattering between layers separated by distance $d$) given by\cite{paco}
\bearr
\tilde{v}(\omega,q) = \frac{2\pi e^2}{\epsilon_0 q}
\frac{\sinh (q d)}
{\sqrt{[\cosh(qd) + {\frac{2\pi e^2}{\epsilon_0 q} } 
\sinh(qd)\chi_0(\omega,q)   ]^2 - 1 }}
\nonumber
\eearr 
and find that the collective mode survives with small quantitative
modifications. 

Let us discuss life time effects, that is beyond RPA.
A remarkable feature of our 
collective modes is that it never enters the particle-hole continuum.
It does not suffer from Landau damping (resonant decay 
into particle-hole pair excitations). To this extent our collective 
modes are sharp and protected; higher order processes will produce
the usual life time broadening, particularly at the high energy end.
However, in real graphite
there are tiny electron and hole pockets in the BZ with a very small 
Fermi energy $\sim 10$ to $20~meV$. This leads to `Landau damping' 
of low energy collective modes around the $\Gamma$ and $K$ points, 
but only in a small momentum region $\Delta k \sim 2k_F \sim 
\frac{1}{50}\frac{\pi}{a} $, where $k_F$ is the mean Fermi momentum of 
the electron and hole pockets. That is, only a few percent of the 
collective mode branch in the entire BZ is Landau damped. 

A small interlayer hopping between neighboring layers 
$t_{\perp}\sim 0.2~eV$ ($<< 2.5~eV$,the in plane hopping matrix element),
has been always invoked in the band theory approaches to understand
various magneto oscillation experiments and also c-axis transport
in graphite. However, a strong renormalization of $t_{\perp}$
is possible, as  anomalously large anisotropic resistivity ratio 
$\frac{\rho_{c}}{\rho_{ab}} \sim 10^4$ have been reported in some early 
experiments on graphite single crystals; a many body renormalization 
is also partly implied by the existence of our spin-1 collective mode 
at low energies.  As the emergence of the small electron and hole 
pockets (cylinders) are due to interlayer hopping, interlayer hopping 
affect the spin-1 collective modes only in a small window of energy 
$0$ and $\sim 0.1~eV$. For the same reason the collective modes do
not have much dispersion along the c-axis. 

Within our RPA analysis the collective mode frequency becomes 
negative at the $\Gamma$ point for $U > U_c \sim 2t$. Because there 
are two atoms per unit cell, this could be either an antiferromagnetic 
or ferromagnetic instability. Other  studies\cite{harigaya,sorella} 
have indicated an AFM instability for $U > U_c \sim 2t$.

Now we discuss the experimental observability of spin-1 collective 
mode branch. The collective mode has a wide energy dispersion 
from $0$ to $\sim 2~eV$. The low energy $ 0$ to $0.05~eV$ part of the 
collective modes determines the nature of the spin susceptibility 
(equation 3) of graphite and leaves its signatures in NMR and ESR 
results. For higher energies we have to use other probes.

Inelastic neutron scattering can be used to study the line shapes 
and dispersion of our spin-1 collective modes. 
However, epithermal neutrons in the energy range $0.1~ eV$ to 
$\sim 1~eV$, rather than the cold and thermal, $0.2$ to $ 50~meV$, 
neutrons is needed in our case, due to the large energy 
dispersion. The dynamic structure 
factor $S({\bf q},\omega)$ as measured by inelastic neutron scattering
is obtained by using our calculated RPA expression for our magnetic 
response function using the relation:
\begin{equation}
        (1-e^{-\beta\omega})~S({\bf q},\omega)=
        -\frac{1}{\pi}~Im~\chi({\bf q},\omega)
\end{equation}
At the present moment one need not concentrate on the energy 
resolution and it will be good to focus on proving the existence
of the spin-1 collective mode by neutron scattering experiments.
As the single phonon density of states of graphite vanish for
energies $> 0.2~eV$, one need not perform spin polarized neutron 
scattering in order to avoid single phonon peaks.

Another probe for studying the spin-1 collective mode is
the spin polarized electron energy loss spectroscopy (SPEELS); 
exchange interaction of the probing electron with the $\pi$-electrons 
of graphite can excite the spin-1 collective mode. As the electron
current and spin depolarization essentially measures the magnetic 
response function, our calculation of $\chi({\bf q},\omega)$ 
(equation 2) can be profitably used to 
interpret the experimental results. 

The square root divergence of density of states at the bottom edge of 
the particle-hole continuum tells us that {\em the low energy spin physics 
is effectively one-dimensional}. To that extent, {\em in a final theory}, 
we may  
expect our spin-1 excitation to be a triplet bound state of `two neutral 
spin$-\frac{1}{2}$ spinons' rather than ` $e^+e^-$ electron-hole
pairs'. Further, as the energy of the spin-1 quantum approaches zero 
the binding energy also approaches zero and the electron-hole bound state 
wave function becomes elliptical, with diverging size. We may then 
view the low energy spin-1 quanta as a `critically
(loosely) bound' two spinon state, very much like the quantum number 
fractionization of the des Cloizeaux-Pearson spin-1 excitation in the  
1d spin-$\frac{1}{2}$ antiferromagnetic Heisenberg model. 
Our result also suggest 
a non-linear sigma model and novel 2 + 1 dimensional bosonization scheme 
for graphite\cite{gb.akbar}. 

We find\cite{gb.nt} that our spin-1 collective mode survives in carbon 
nano-tubes in a modified fashion. Preliminary study shows 
that three dimensional semimetals Bi, HgTe and $\alpha$-Sn do not 
have spin-1 collective modes at low energies, because of quadratic 
dispersion at the zero gap. 

G.B. thanks B.S. Shastry for a discussion. S.A.J. thanks 
the Institute of Mathematical Sciences, Madras for a Visiting Student 
Fellowship and Professor H. Arfaei of Sharif University of Technology, 
Tehran for his continued support.

\end{multicols}

\end{document}